\documentclass[twocolumn]{aastex62}
\usepackage{amsmath}
\pdfoutput=1

\submitjournal{ApJ}
\shorttitle{EUV late phase}
\shortauthors{Zhang et al.}

\begin{document}

\title{Implications for additional plasma heating driving the extreme-ultraviolet late phase of a solar flare with microwave imaging spectroscopy}

\correspondingauthor{Hui Tian}
\email{huitian@pku.edu.cn}

\author{Jiale Zhang}
\affiliation{School of Earth and Space Sciences, Peking University, Beijing 100871, China}

\author[0000-0002-0660-3350]{Bin Chen}
\affiliation{Center for Solar-Terrestrial Research, New Jersey Institute of Technology, Newark, NJ 07102, USA}

\author[0000-0003-2872-2614]{Sijie Yu}
\affiliation{Center for Solar-Terrestrial Research, New Jersey Institute of Technology, Newark, NJ 07102, USA}

\author{Hui Tian}
\affiliation{School of Earth and Space Sciences, Peking University, Beijing 100871, China}
\affiliation{Key Laboratory of Solar Activity, National Astronomical Observatories, Chinese Academy of \\
	Sciences, Beijing 100012, China}

\author{Yuqian Wei}
\affiliation{Center for Solar-Terrestrial Research, New Jersey Institute of Technology, Newark, NJ 07102, USA}

\author{Hechao Chen}
\affiliation{School of Earth and Space Sciences, Peking University, Beijing 100871, China}

\author{Guangyu Tan}
\affiliation{School of Earth and Space Sciences, Peking University, Beijing 100871, China}

\author{Yingjie Luo}
\affiliation{Center for Solar-Terrestrial Research, New Jersey Institute of Technology, Newark, NJ 07102, USA}

\author{Xingyao Chen}

\affiliation{Key Laboratory of Solar Activity, National Astronomical Observatories, Chinese Academy of \\
	Sciences, Beijing 100012, China}



\begin{abstract}

Extreme-ultraviolet late phase (ELP) refers to the second extreme-ultraviolet (EUV) radiation enhancement observed in certain solar flares, which usually occurs tens of minutes to several hours after the peak of soft X-ray emission. The coronal loop system that hosts the ELP emission is often different from the main flaring arcade, and the enhanced EUV emission therein may imply an additional heating process. However, the origin of the ELP remains rather unclear. Here we present the analysis of a C1.4 flare that features such an ELP, which is also observed in microwave wavelengths by the Expanded Owens Valley Solar Array (EOVSA). Similar to the case of the ELP, we find a gradual microwave enhancement that occurs about three minutes after the main impulsive phase microwave peaks. Radio sources coincide with both footpoints of the ELP loops and spectral fits on the time-varying microwave spectra demonstrate a clear deviation of the electron distribution from the Maxwellian case, which could result from injected nonthermal electrons or nonuniform heating to the footpoint plasma. We further point out that the delayed microwave enhancement suggests the presence of an additional heating process, which could be responsible for the evaporation of heated plasma that fills the ELP loops, producing the prolonged ELP emission.

\end{abstract}


\section{Introduction} \label{sec:intro}

Solar flares are explosive energy release events powered by magnetic reconnections in the solar corona \citep{2011SSRv..159...19F,2017LRSP...14....2B}. They are usually associated with enhanced electromagnetic radiation over a wide range of wavelengths, from radio to gamma-rays. The evolution of a solar flare includes two major phases, the impulsive phase and the gradual phase. Most of the stored magnetic energy is released in the impulsive phase, which is further converted to plasma heating, particle acceleration, and bulk flows. In many cases, a dramatic increase in hard X-ray (HXR) emitted by precipitated nonthermal electrons appears first, followed by an enhancement in soft X-ray (SXR) and extreme-ultraviolet (EUV) emission from the heated plasma. Cooling of the post-flare loops is dominant in the gradual phase and solar radiation eventually returns to its original state. The SXR emission peak is usually used to demarcate the impulsive phase from the gradual phase: A sharp increase in the SXR flux refers to the impulsive phase and the subsequent slow decay represents the latter. 

The Atmospheric Imaging Assembly \citep[AIA;][]{2012SoPh..275...17L} and the EUV Variability Experiment \citep[EVE;][]{2012SoPh..275..115W} onboard the Solar Dynamics Observatory \citep[SDO;][]{2012SoPh..275....3P} have provided high-quality solar EUV images and spectra, respectively, to bring new insights into the complex dynamics of solar flares in recent years. \cite{2011ApJ...739...59W} examined EVE flare observations and found that some flares exhibit a second peak in warm EUV emission lines (e.g., Fe~{\sc{xvi}}~335\AA\, with a formation temperature of $\sim$3 MK) tens of minutes or several hours after the peak of soft X-ray radiation, which is referred to as the EUV late phase (ELP). The multi-band AIA images reveal that the ELP radiation originates from loops residing in the same active region (AR) but much larger and taller than the original flaring loops, raising questions on the physical connection between these two sets of loops. The unique variation of EUV irradiance during the ELP has also drawn great interest owing to its potential geoeffectiveness. Enhanced EUV irradiance from the Sun could cause thermospheric heating and ionospheric disturbances \citep{1971ApJ...164..151K} in the Earth's upper atmosphere, leading to a stronger atmospheric drag on low-earth-orbit satellites \citep{1989JSpRo..26..439W} and interruption of radio communications \citep{2009RaSc...44.0A17T}.

Through a decade of investigations, much progress has been made in understanding the causes of delayed EUV emission peaks in ELP events. The additional heating scenario was first proposed, suggesting a long-term, mild energy release process in the flare gradual phase or the EUV late phase afterwards which directly warms up the ELP loops to intermediate temperature \citep{2011ApJ...739...59W,2012arXiv1202.4819H,2013ApJ...773L..21D,2017ApJ...835....6K,2019ApJ...878...46Z}. An alternative explanation is known as the extended cooling process. As previous works suggested that the cooling timescale of coronal loops increases with greater loop length \citep[e.g.][]{1995ApJ...439.1034C}, some argued that the higher ELP loops are heated almost simultaneously with the lower loops but cool down much slower, resulting in prolonged EUV emission \citep{2013ApJ...768..150L,2017A&A...604A..76M,2018ApJ...863..124D,2020ApJ...890..158C}. It has also been pointed out that the additional heating and extended cooling can be both at work in some ELP events \citep{2013ApJ...778..139S,2014ApJ...793...85L}, indicating that the two mechanisms are not mutually exclusive.

Using a 0-D hydrodynamic loop model called Enthalpy-Based Thermal Evolution of Loops \citep[EBTEL;][]{2008ApJ...682.1351K,2012ApJ...752..161C,2016ApJ...829...31B}, and considering a time-varying heating function and a continuous cooling process, one can synthesize light curves of solar flares with ELPs that match well with observations \citep{2012arXiv1202.4819H,2014ApJ...793...85L,2014ApJ...781..120L,2018ApJ...863..124D,2019ApJ...878...46Z}. However, the true nature of the heating source of ELP loops is not revealed. Some researchers suspect that the nonthermal electron beams are responsible for the heating of the ELP loops, but their presence and role remains unclear.
\cite{2021A&A...650A..88Z} recently investigates the energy partition in a flare with an ELP, and finds that the energy of flare-accelerated electrons is sufficient to heat both the main flaring loops and the ELP loops. It is widely acknowledged that X-ray and microwave observations are highly complementary to each other in tracing and quantifying nonthermal electrons in solar flares \citep{2011SSRv..159..225W}. If nonthermal electrons are the important source of energy that drives the ELP,  observations of X-ray and microwave should reveal key information on the origin of the ELP.

The Expanded Owens Valley Solar Array \citep[EOVSA,][]{2016JAI.....541009N} is capable of providing imaging spectroscopy of the full Sun over a broad spectral range (1--18 GHz), which is suited to study a wide range of topics from solar flares, active regions, to the quiet Sun. The EOVSA observations have greatly advanced our understanding of erupting flux ropes \citep{2020ApJ...895L..50C}, flare current sheets \citep{2020NatAs...4.1140C} and flare arcades/loop-top regions \citep{2018ApJ...863...83G,2020Sci...367..278F,2020ApJ...900...17Y,2020ApJ...905..165R,2021ApJ...908L..55C}. Here we present results from the first joint analysis of SDO/AIA and EOVSA observations of a confined flare that features an ELP. After a brief overview of the flare in Section \ref{sec:observations}, we discuss the ELP and its relation to the microwave emission in Section \ref{sec:ELP}. We further obtain the physical properties of the radio source using spatially resolved microwave spectra in Section \ref{sec:fittings}, and discuss the cause of ELP in Section \ref{sec:discussions}. Finally, we summarize our results in Section \ref{sec:conclusions}.

\section{EOVSA observations} \label{sec:observations}
\begin{figure*}
	\centering
	\includegraphics[width=\linewidth]{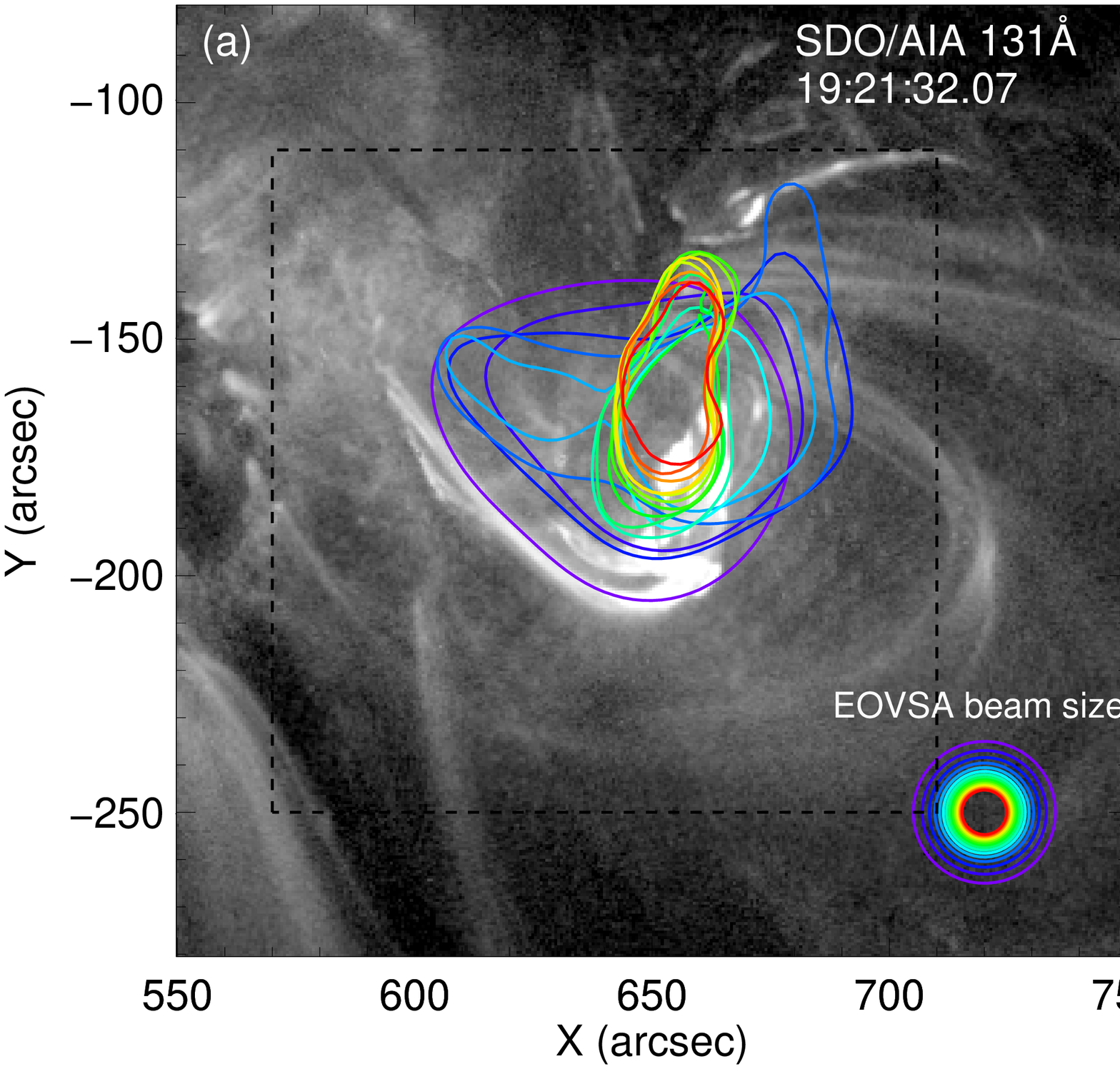}
	\caption{(a) Multi-frequency radio emission contours (30\% of the maximum) overplotted on an AIA 131 \AA\ image. The sizes of the restoring beams are shown in the bottom right corner. The dotted black square marks the field of view in Figures~\ref{fig:figure3} and \ref{fig:figure4}. (b) Same as (a) but for a different time. The dotted red and blue rectangles mark region A and region B, respectively. The solid squares with a size of 16''$\times$16'' within the two regions mark box A and box B, which are chosen for spectral fittings in Figure \ref{fig:figure8}.\label{fig:figure1}}
\end{figure*}

The C1.4 flare starts around 19:21 UT on July 14th, 2017 in NOAA AR 12665 (SOL2017-07-14T19:21), which is well observed by SDO/AIA and EOVSA. In the flare impulsive phase, a high-frequency ($>$6 GHz) microwave source shows up near the bright and compact flaring loops in AIA 131 \AA\ images (Figure \ref{fig:figure1}(a)). As the flare approaches its gradual phase, a set of more extended outer loops (we will later 
refer to them as the ELP loops) begin to brighten up. At the eastern footpoint, a new radio source emerges. To distinguish it from the previous one, we refer to the primary and secondary sources as regions A and B, respectively. The two radio sources coincide with both footpoints of the new outer loops.

\begin{figure*}
	\centering
	\includegraphics[width=\linewidth]{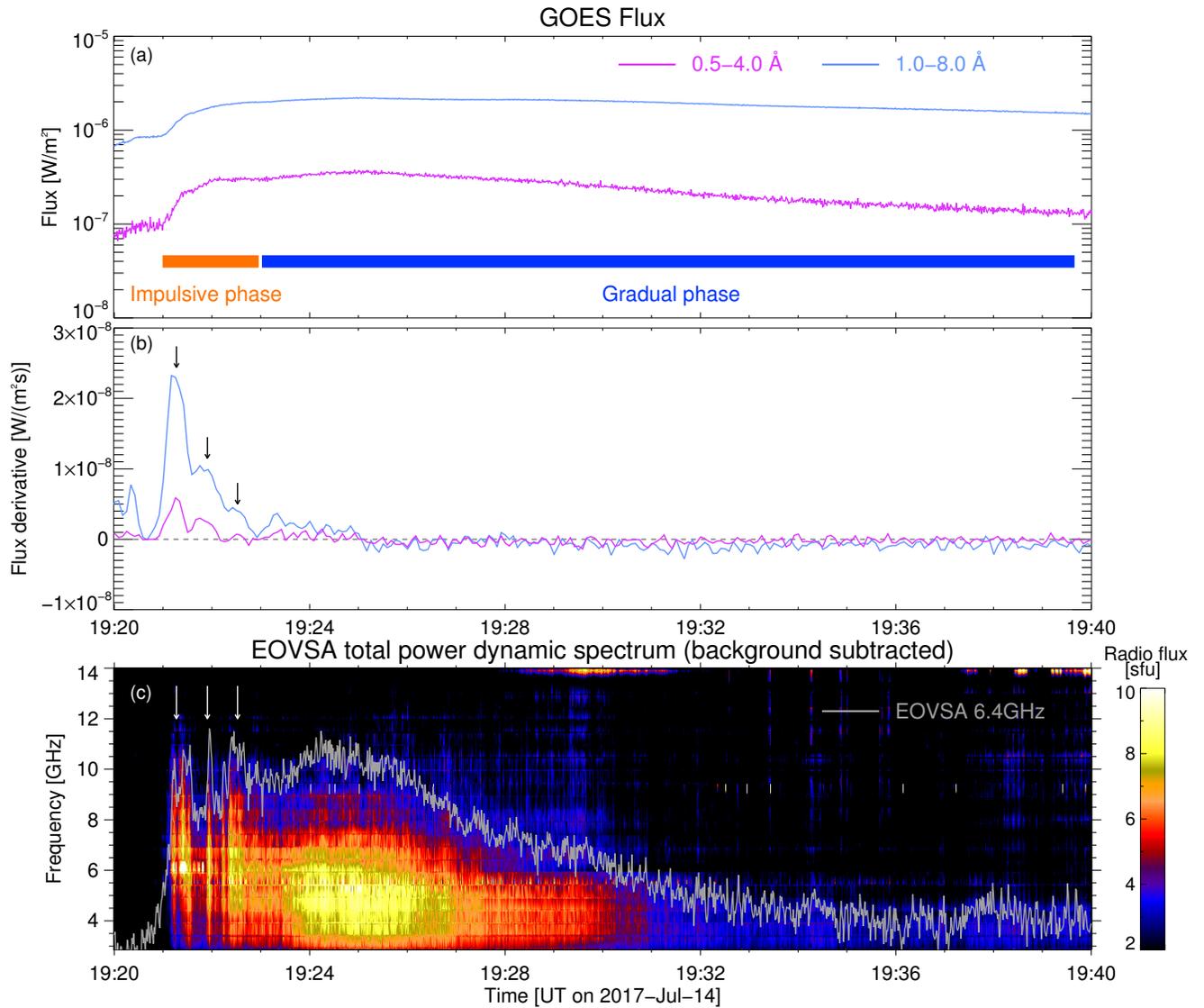}
	\caption{(a) GOES SXR light curves of this event. The pink and blue curves stand for the wavelength ranges of 0.5--4.0 \AA\ and 1.0--8.0 \AA\ . The periods of the impulsive phase and the gradual phase are marked by the thick orange and blue bars, respectively. (b) The time derivative of GOES 1.0--8.0 \AA\ and 0.5--4.0 \AA, after a 5 s averaging on the 1 s cadence data. The arrows mark three main peaks. (c) EOVSA total-power microwave dynamic spectrum. A pre-flare background subtraction is performed (the spectra at 19:20:00.50 UT is selected as the background). The temporal evolution of the EOVSA 6.4 GHz flux is overplotted. Three emission peaks are marked by the arrows and compared with the peaks in (b).\label{fig:figure2}}
\end{figure*}

In Figure \ref{fig:figure2}(a), we compare the SXR light curves from the Geostationary
Operational Environment Satellite (GOES) with the EOVSA microwave total-power dynamic spectrum. The flare impulsive phase lasts from 19:21 UT to 19:23 UT and the gradual phase follows.  In the microwave total-power dynamic spectrum (Figure \ref{fig:figure2}(c)), we can find multiple short-duration microwave bursts in the flare impulsive phase. Three main groups of microwave bursts can be identified (arrows in Figure \ref{fig:figure2}(c)), the timing of which corresponds very well to the peaks in the derivative of the SXR light curve. This feature is well known as the Neupert effect \citep{1968ApJ...153L..59N} characteristic of the flare impulsive phase. These microwave and SXR derivative peaks may suggest the intermittent manner of the energy release and particle acceleration processes during the flare impulsive phase \citep{2004ApJ...605L..77A,2009ApJ...694L..74N,2016ApJ...828..103T,2018ApJ...866...64C}. At the end of the flare impulsive phase around 19:23 UT, the impulsive microwave bursts also diminish. Meanwhile, another microwave emission component develops in the dynamic spectrum. Different from the short-duration microwave bursts during the impulsive phase, the delayed microwave emission at the beginning of the gradual phase has a much longer duration ($\sim$7 min; peaking at $\sim$19:24 UT). It also features a much smoother behavior in the dynamic spectrum with a slow variation in time. This striking difference implies that the emission in the two stages arises from different energy release processes. 

\section{The EUV late phase} \label{sec:ELP}
\begin{figure*}
	\centering
	\includegraphics[width=\linewidth]{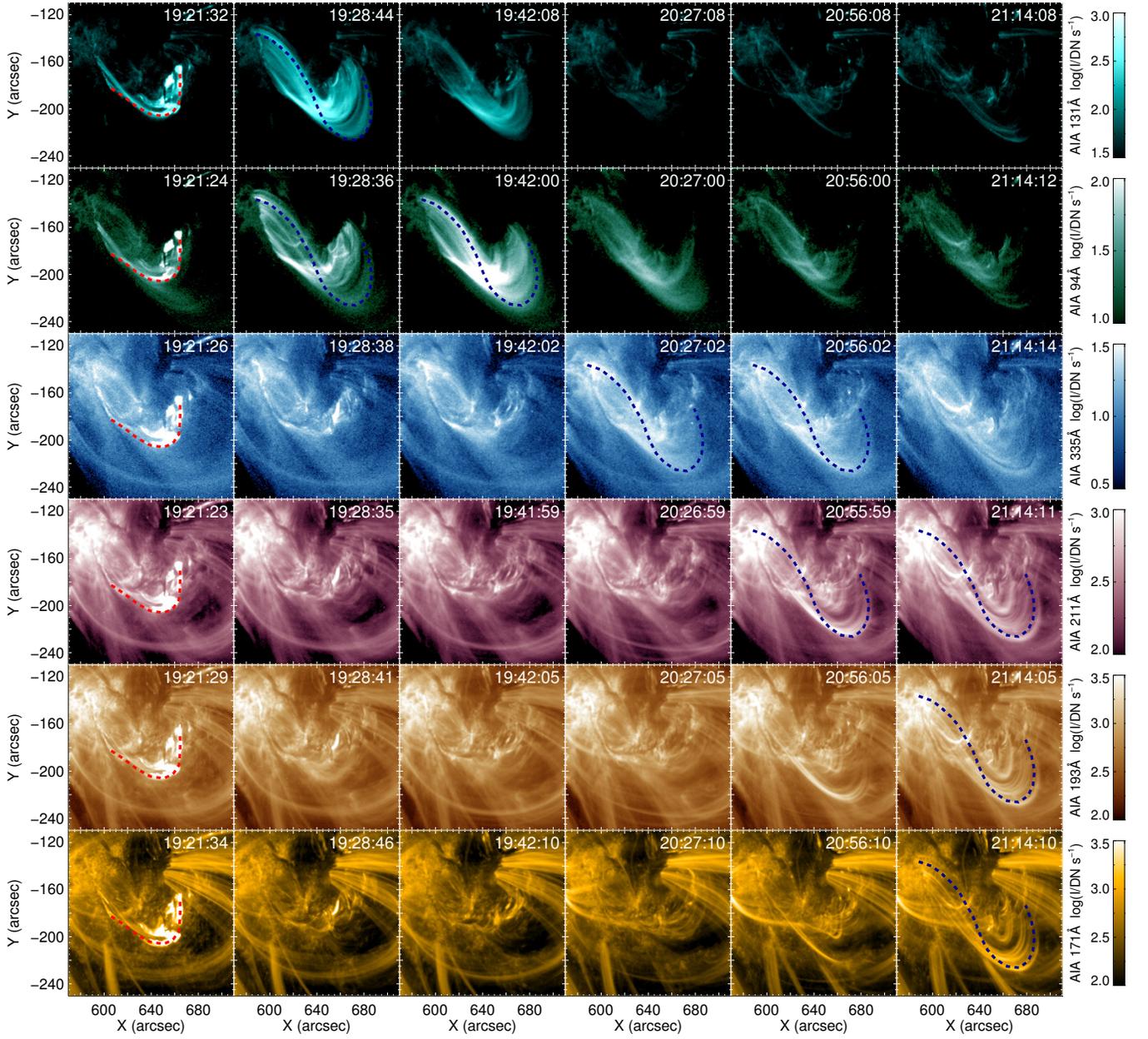}
	\caption{Image sequences of AIA 131 \AA, 94 \AA, 335 \AA, 211 \AA, 193 \AA, and 171 \AA. The peak temperature of the corresponding response function decreases from top to bottom. The first column shows images taken in the flare impulsive phase and the other columns show images taken in the EUV late phase. The red (blue) dotted line outlines the shape of the original flaring loops (ELP loops).\label{fig:figure3}}
\end{figure*}

The EUV late phase of the flare starts around 20:00 UT, when AIA 335 \AA\ passband shows another episode of enhanced emission. In AIA 335 \AA\ images (the third row in Figure \ref{fig:figure3}), a set of higher-lying loops brighten up above the original flaring loops. During that time, the GOES SXR flux almost returns to its original state and AIA 131 \AA\ emission in the area is very faint, which meets the criteria for ELP described in \cite{2011ApJ...739...59W}. We notice that the post-flare loops appearing in the flare gradual phase (19:23UT - 20:00UT) in AIA 131 \AA\ and AIA 94 \AA\ images share a similar morphology with the ELP loops later in AIA 335 \AA\ images, indicating that they are likely the same loops. Although the original definition of ELP loops in \cite{2011ApJ...739...59W} refers to those seen in medium-temperature EUV filters, some researchers have argued that the ELP loops in some cases could be firstly heated up to $\sim$10 MK and appear earlier in high-temperature AIA passbands \citep{2013ApJ...778..139S,2018ApJ...863..124D}. Following this line of argument and considering the same morphology of the high-lying loops seen in AIA 131 \AA, 94 \AA, and 335 \AA, we refer to these loops as “ELP loops". The ELP loops are also seen in AIA 211 \AA\, 193 \AA\, and 171 \AA\ images at later times (after $\sim$21:00 UT) before finally fading away. The sequentially delayed brightening of the loops in EUV bands that have a sensitivity to cooler and cooler coronal plasma (from $\sim$10 MK for AIA 131 \AA\ to $\sim$1 MK for AIA 171 \AA; \citealt{2010A&A...521A..21O}) is consistent with the typical ELP phenomena \citep{2011ApJ...739...59W}. It is worth noting that AIA 131 \AA\ emission is also contributed by cold Fe~{\sc{viii}}~ lines \citep{2010A&A...521A..21O} and there is a chance that the cold emission is confused with the hot one. We do not consider it a problem here, as the AIA 131 \AA\ emission takes on entirely different characteristics from those cold EUV filters (for instance AIA 193 \AA\ and 171 \AA) during the gradual phase (see the second and third columns in Figure \ref{fig:figure3}).

\begin{figure*}
	\centering
	\includegraphics[width=\linewidth]{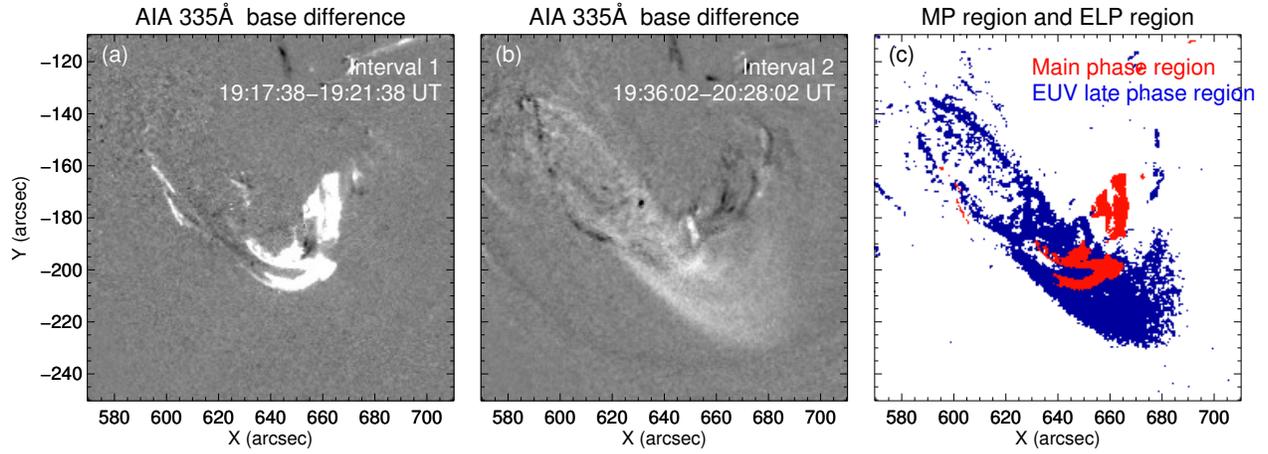}
	\caption{(a) and (b) Base-difference images of AIA 335 \AA\ for the main impulsive phase and the EUV late phase, respectively. The times used for producing the difference images are indicated in each panel. (c) The main phase region and the ELP region are marked in red and blue, respectively. The two regions are made up of selected pixels from images in panel (a) and panel (b) using two different intensity thresholds. The threshold for the main phase region is 18 DN/s and that for the ELP region is 4 DN/s. Pixels initially classified to the main phase region are excluded from the EUV late phase region.\label{fig:figure4}}
\end{figure*}

\begin{figure*}
	\centering
	\includegraphics[width=0.7\linewidth]{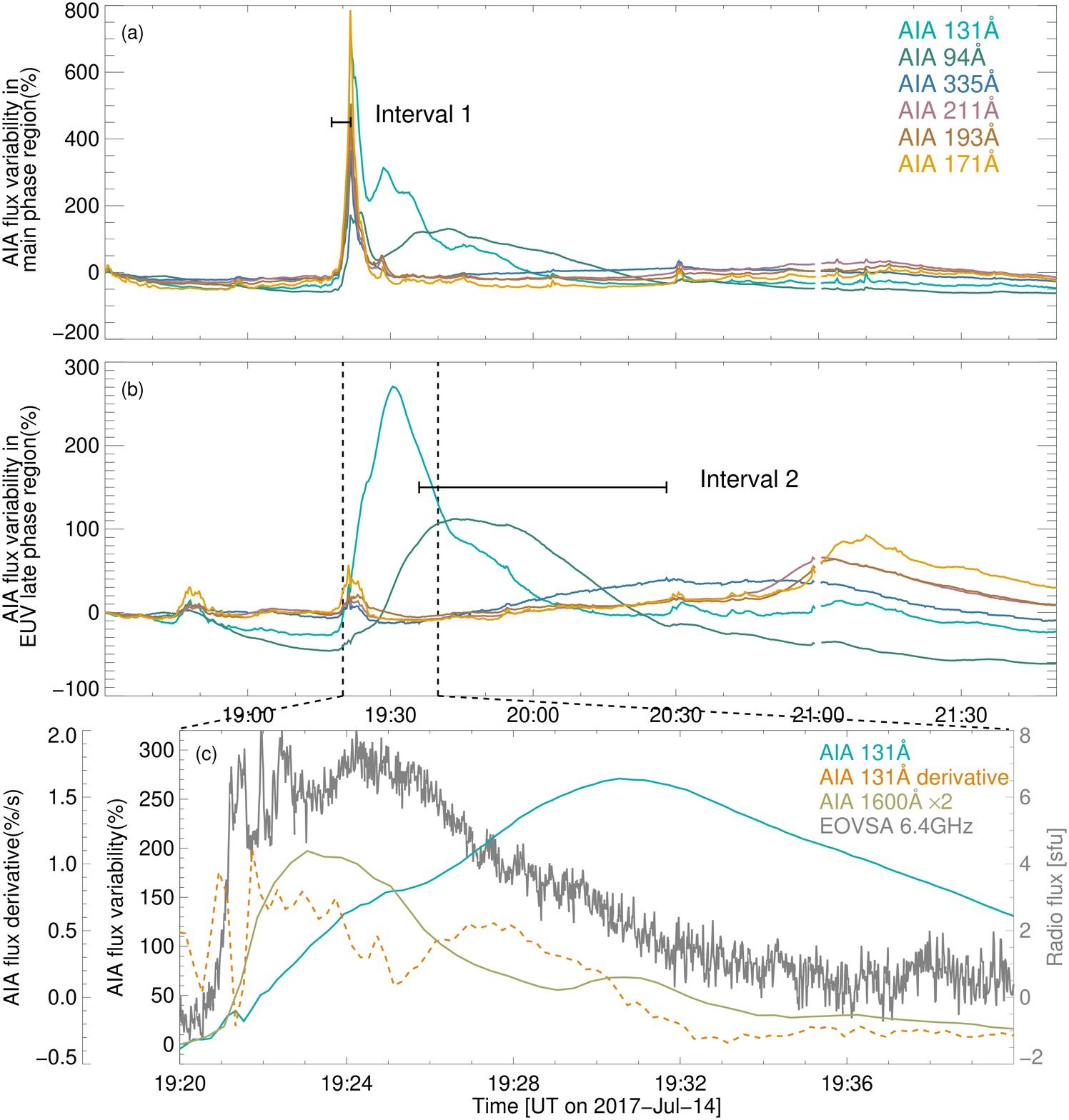}
	\caption{(a) Light curves of AIA 131 \AA, 94 \AA, 335 \AA, 211 \AA, 193 \AA, 171 \AA in the main phase region (the red region in Figure \ref{fig:figure4}(c)). The flux variability is defined by $(I_{time}-I_{ref})/I_{ref}\times100\%$ where $I_{time}$ and $I_{ref}$ refer to the total intensities at a certain time and a reference time (18:30 UT), respectively. The range of the interval 1 is marked in the chart. (b) Light curves of AIA 131 \AA, 94 \AA, 335 \AA, 211 \AA, 193 \AA, 171 \AA in the ELP region (the blue region in Figure \ref{fig:figure4}(c)). The range of the interval 2 is marked in the chart. (c) Light curve of AIA 131 \AA in the ELP region, its derivative, light curve of AIA 1600 \AA of the localized brightenings near the ELP loop eastern footpoint (integrated within the blue contours in the first column of Figure \ref{fig:figure7}(a)) and the temporal evolution of EOVSA 6.4 GHz (from the background-subtracted total power dynamic spectrum). The light curve of AIA 1600 \AA is multiplied by a factor of 2 to aid visual comparison.\label{fig:figure5}}
\end{figure*}

By applying different intensity thresholds on two selected base-difference images (see Figures \ref{fig:figure4}(a) and (b)), we identify the pixels of the main flaring loops and the ELP loops, shown as red and blue color, respectively, in Figure \ref{fig:figure4}(c). It is clear that the main flare phase occurs in a set of relatively compact loops, but the ELP appears to happen in an extended higher-lying loop system. To better demonstrate the temporal evolution of  EUV emission in these two regions, we obtain the light curves of six AIA passbands in the two regions by summing up the total intensity of all the identified pixels. Figure \ref{fig:figure5}(a) reveals that the light curves of different AIA passbands in the main flaring region simultaneously peak during the impulsive phase around 19:22 UT. However, a different behavior is present in the integrated EUV light curves over the ELP region (Figure \ref{fig:figure5}(b)). Emission in high-temperature AIA passbands like AIA 131 \AA\ and 94 \AA\ rises shortly after the first emission peak in the main phase region, followed by AIA 335 \AA\ and later by 211 \AA\, 193 \AA\, and 171 \AA. The sequentially delayed emission peaks reflect an extended cooling process (e.g. \citealt{2012ApJ...753...35V,2022ApJ...928...98H}) that lasts from 19:25 UT to 21:30 UT, during which the heated ELP loops cool down to the corresponding characteristic temperatures of different AIA passbands at different times. Among all, the AIA 335 \AA\ passband shows the longest emission enhancement, almost covering the entire 2-hour-long ELP. 

\begin{figure*}
	\centering
	\includegraphics[width=\linewidth]{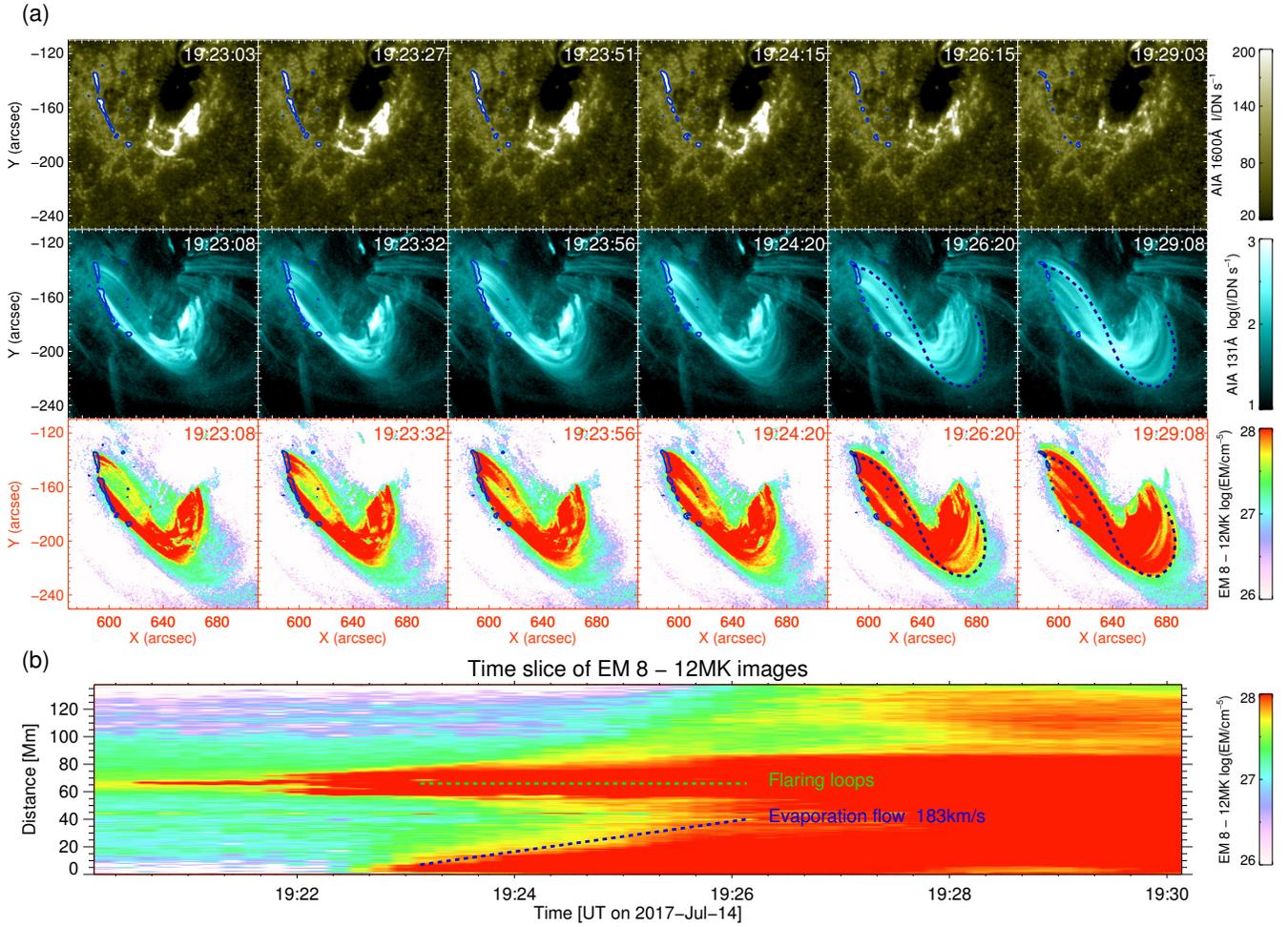}
	\caption{(a) First to third row: sequences of AIA 1600 \AA, AIA 131 \AA, and EM image series in the temperature range of 8--12 MK, respectively. The blue contours mark the localized brightenings in AIA 1600 \AA, around the eastern footpoint of the ELP loops. The blue dotted lines indicate the shape of the ELP loops. (b) Time-distance diagram of EM 8--12 MK images obtained along the blue dotted line in (a). The location of the flaring loops and the motion of the evaporation flow from the eastern footpoint are marked by the green and blue dotted lines, respectively. \label{fig:figure6}}
\end{figure*}

Similar to the delayed microwave enhancement discussed in Section \ref{sec:observations}, the EUV enhancement in AIA 131 \AA\ also occurs shortly after the impulsive phase. In Figure \ref{fig:figure5}(c), the time profile of AIA 131 \AA\ in the ELP region and its derivative are compared with the microwave light curve. The emission of AIA 131 \AA\ undergoes a continuous increase as the delayed microwave emission rises and decays, and reaches its peak $\sim$6 minutes after the peak of microwave light curve. We also produce the derivative of the AIA 131 \AA\ light curve of the ELP region, shown as the orange dashed curve in Figure \ref{fig:figure5}(c). The AIA 131 \AA\ derivative resembles the characteristics of the microwave 6.4 GHz light curve, despite some subtle differences. 
Because the AIA 131 \AA\ passband is sensitive to the flare-heated plasma at a similar temperature to the GOES 1--8 \AA\ band, similar to the main impulsive phase, the similarity of the AIA derivative of the ELP region and the delayed microwave light curve is also suggestive of the Neupert effect at play, which is likely responsible for the heating of the ELP loops. We also find clear indication of chromospheric heating during the flare gradual phase in the AIA 1600 \AA\ images (Figure \ref{fig:figure6}(a)), which is presented as several localized brightenings at the eastern footpoint and patches of brightenings at the western footpoint.
We derive the AIA 1600 \AA\ ultraviolet(UV) light curve by integrating the brightening regions near the eastern footpoints of ELP loops (indicated by the blue contours in Figure \ref{fig:figure6}(a)). 
The eastern footpoints are chosen because they are far away from the main flaring region toward the west, which minimize the contribution from those not associated with the ELP. 
The enhanced AIA 1600 \AA\ emission lasts from 19:22 UT to 19:28 UT, which indicates a lasting chromospheric heating process. In general, both the enhanced microwave emission and the UV ribbon/footpoint brightenings during solar flares are good indicators of the energy deposition to the loop footpoints \citep{2015ApJ...811..139T,2017ApJ...841L...9L,2019ApJ...870..109Z}. The close temporal relationships among the EUV 131 \AA\ derivative, UV footpoint emission, and the microwave emission suggest that the energy input at the loop footpoints likely drives the heating of the ELP loops through chromospheric evaporation.

In addition to the temporal association, moreover, we find that the EUV enhancement in the ELP loops is spatially associated with the microwave enhancement. In Figure \ref{fig:figure1}(b), the two high-frequency ($>$6 GHz) radio sources coincide with the primary flaring region (or the western footpoint of the ELP loop system) and the conjugate, eastern footpoint of the ELP loop system where the UV footpoint emission is also presented. It further supports the chromospheric evaporation process due to footpoint heating. As there is no obvious flaring activity near the eastern footpoint (or region B), we suspect that the deposited energy in region B is somehow transported from the flaring region (or region A). The energy transfer could be realized by thermal conduction or nonthermal electron beams in the realm of solar flares, when energy is carried by thermal or nonthermal particles, respectively. These transported and finally precipitated particles heat the local chromospheric plasma, which may contribute to the heating of ELP loops through a process similar to the chromospheric evaporation in the main flare impulsive phase, but perhaps in a much more gentle and gradual fashion \citep{1985ApJ...289..425F}.

To verify our scenario, we take a close look at the formation process of the ELP loops. In AIA 131 \AA\ images (second row of Figure \ref{fig:figure6}), apart from the bright flaring loops, a plume of evaporated hot plasma rises from the eastern footpoint and fills in the ELP region. 
To study the thermodynamics of the ELP loops, we adopt the modified version of the sparse inversion code \citep{2015ApJ...807..143C,2018ApJ...856L..17S} using images of all AIA EUV channels except for the 304 \AA. This code is known to be well suited for investigations of hot plasma dynamics during flares (e.g. \citealt{2020ApJ...898...88X,2021Innov...200083S}). The emission measure (EM) maps in the temperature range of 8--12 MK lead to the same conclusion: Evaporation flow propagates upward from the area with chromospheric brightenings and warms up the ELP loops to more than 10 MK. The time slice of the EM images reveals the motion of the evaporation flow from the eastern footpoint, which gives a speed of $\sim$183 km s$^{-1}$ in projection (note that the true velocity should be higher given the projection effect). Evaporation flow from the western footpoint is relatively inapparent, presumably due to the covering of the flaring loops. But its heating effect should not be neglected. The process is well consistent with the chromospheric evaporation scenario \citep{2014ApJ...797L..14T,2015ApJ...811..139T,2015ApJ...807L..22G,2015ApJ...811....7L,2015ApJ...813...59L,2017ApJ...841L...9L,2019ApJ...870..109Z}, supporting that the evaporated plasma from the footpoints is one of the important energy sources for the heating of ELP loops. By saying so, we cannot exclude the direct conductive heating from the flaring loops to the ELP loops, although our data cannot determine its exact role in this process. 

\section{Microwave Spectral Analysis} \label{sec:fittings}
\begin{figure*}
	\centering
	\includegraphics[width=\linewidth]{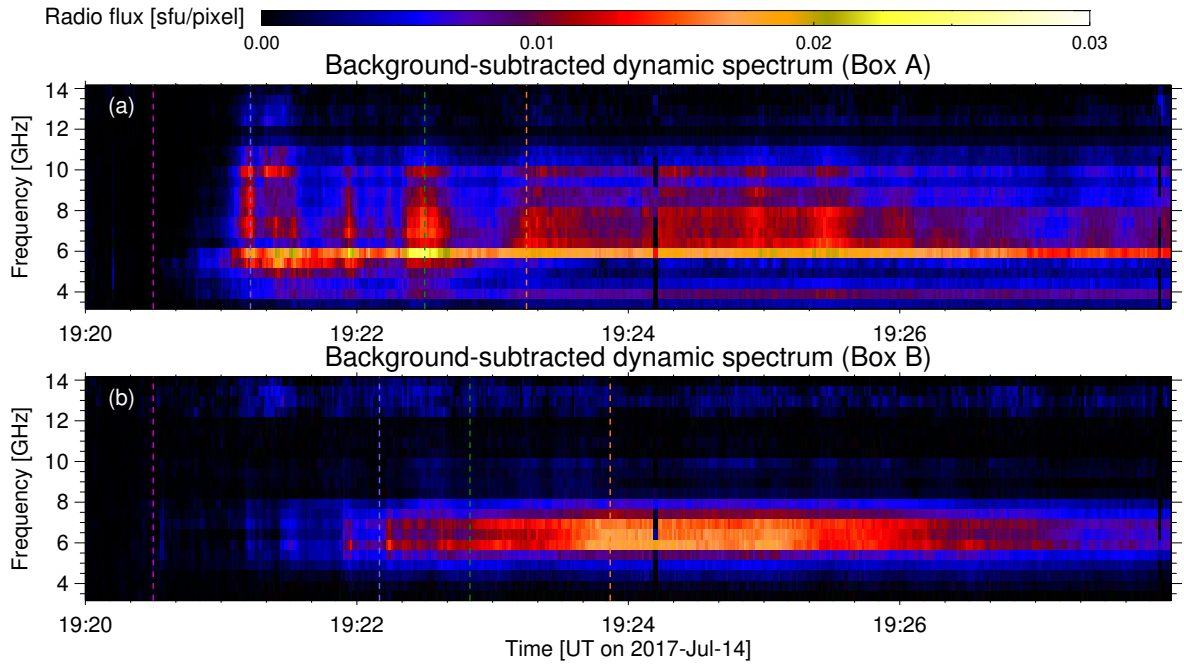}
	\caption{(a,b) The background-subtracted dynamic spectra for box A and box B. The background time is 19:20:00.50 UT. Colored lines indicate the times for which spectral fitting is performed.\label{fig:figure7}}
\end{figure*}

In Section \ref{sec:ELP}, we speculate that the footpoint heating in region B is closely related to that in region A. To better compare these two regions, we obtain the dynamic spectra (Figure \ref{fig:figure7}) of the two boxes (box A and box B in Figure \ref{fig:figure1}(b)) as representatives, which are also located at both ELP loop footpoints. It is evident that the microwave emission is mainly enhanced in region A in a bursty manner during the impulsive phase, when the flaring activity and the brightenings of the flaring loops occur. When it comes to the gradual phase (after 19:23 UT),  the two regions share a similar emission pattern in the dynamic spectra, a 7-minute-long continuous microwave enhancement. It suggests that the delayed heating in the two regions could originate from the same energy release and transfer process.

\begin{figure*}
	\centering
	\includegraphics[width=\linewidth]{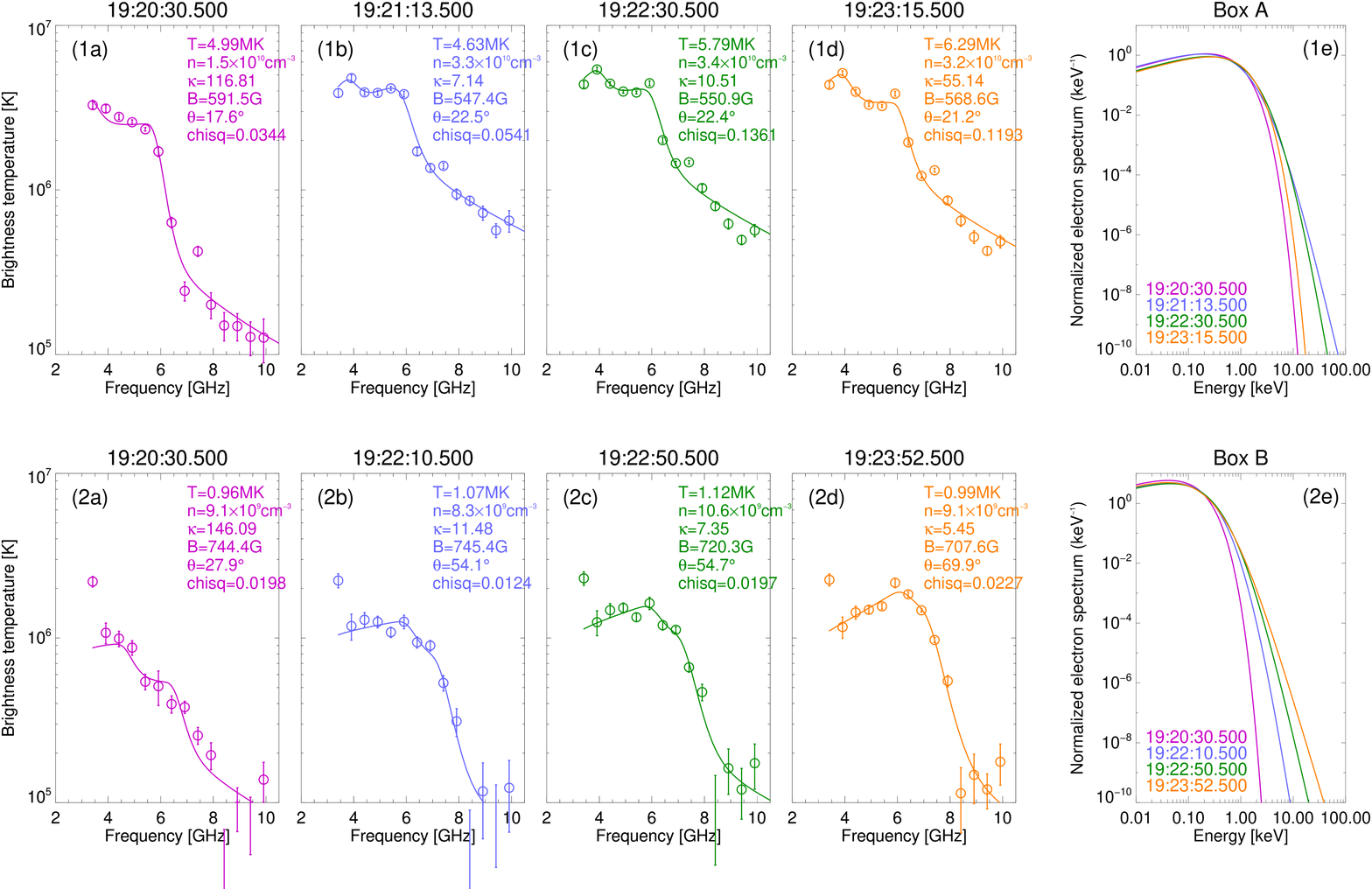}
	\caption{(1a$-$1d, 2a$-$2d) Microwave spectra in the box A and box B, respectively. The circle symbols (error bars) show the data points (uncertainties). The data points at 3.4 GHz for spectra in box B are not taken into account in spectral fittings. Colored solid lines refer to the corresponding best-fit results in our least-square-based fitting. Fitting parameters are listed on the right ($T$, background plasma temperature [MK]; $n$, background electron density [$\rm cm^{-3}$]; $\kappa$, the kappa index in kappa distribution; $B$, the magnetic field [G]; $\theta$, the viewing angle relative to the magnetic field [$^{\circ}$]; chisq, the chi-square of the fitting results.) (1e, 2e) The normalized electron spectra obtained from the best-fit electron distribution function. The colored lines refer to the electron energy distributions at the four times.\label{fig:figure8}}
\end{figure*}

The microwave emission spectrum is dependent on the magnetic field, electron energy, ambient plasma density and temperature, among others, thus providing a diagnostic tool to determine the local plasma properties in the microwave source (see \citealt{1985ARA&A..23..169D,1998ARA&A..36..131B} for a review). We use box A and box B to carry out spectral fittings for the two regions. In Figure~\ref{fig:figure8}, we show the microwave spectra obtained within the small boxes in each frequency band at selected times. The uncertainty of each data point is mainly determined by the root-mean-square value of fluctuations in a region far away from the microwave source. We also take into account the systematic uncertainty in each frequency band introduced by the frequency-dependent spatial resolution of the EOVSA instrument \citep{2013SoPh..288..549G,2020Sci...367..278F}. We use the fast gyrosynchrotron code \citep{2010ApJ...721.1127F} to calculate the microwave spectra. The microwave spectra reveal a steep fall-off above peak frequency and resemble the shape of the thermal gyroresonance spectrum\citep{1985ARA&A..23..169D}. However, we find that the rising radio flux below peak frequency (especially in box B) differs from the thermal case, which usually gives a rather flat spectrum in the optically thick part. So we use kappa distribution as the model for electron distribution.
Kappa distribution describes a smooth transition from the thermal distribution to a nonthermal power-law tail and has been successively applied in HXR \citep{2009A&A...497L..13K,2013ApJ...764....6O} and microwave \citep{2015ApJ...802..122F} spectral diagnostics.

We treat all the input parameters as free parameters to perform the spectral fitting (the high-energy cutoff is fixed at $E_{\rm max}=1 \rm\; MeV$). As shown in Figure \ref{fig:figure8}, the temperature and electron density reveal an increase with time in box A, with no significant variation in box B. Box B does show some changes in magnetic field strength and angle, which might be explained by the change of the optically thick region in the microwave domain \citep{2015ApJ...802..122F}. Most notably, we find a good correlation between the peak radio flux ($\sim$6 GHz) and the kappa index. Especially for box B, the rising peak flux corresponds to the decreasing kappa index. We find a minimum kappa index of 7.14 at box A and that of 5.45 at box B. To evaluate the fitting results, following \cite{2020NatAs...4.1140C}, we also employ the Markov chain Monte Carlo (MCMC) analysis, implemented by an open-source Python package {\tt emcee} \citep{2013PASP..125..306F} and shown in the form of corner plot (Figures \ref{fig:figure10} and \ref{fig:figure11}). The minimization-based fitting and MCMC analysis yield very similar results, indicated by the green and orange lines, respectively, in the corner plots.

In Figures \ref{fig:figure8}(1e) and (2e), we show the best-fit electron energy spectra of the four selected times normalized to an electron density of unity. The normalized energy distribution function $F_{\kappa}(E)\;(\rm keV^{-1})$ in Kappa distribution can be written as \citep{1991PhFlB...3.1835S,2009A&A...497L..13K,2013ApJ...764....6O}

\begin{equation}
	\begin{split}
		F_{\kappa}(E)=\frac{2\sqrt{E}}{\sqrt{\pi(k_BT_{\kappa})^3}}\frac{\Gamma(\kappa+1)}{(\kappa-3/2)^{3/2}\Gamma(\kappa-1/2)} \\ \times\left[1+\frac{E}{k_BT_{\kappa}(\kappa-3/2)}\right]^{-(\kappa+1)},
	\end{split}
	\label{equ:equ2}
\end{equation}
where $E$ is the particle energy, $\kappa$ is the kappa index, $k_B$ is the Boltzmann constant, $\Gamma$ is the Gamma function. $T_{\kappa}$ is the kappa temperature which is related to the temperature $T_M$ derived from the spectral fitting,  $T_{\kappa}=T_M[\kappa/(\kappa-3/2)]$. In both Boxes A and B, the hardening electron spectrum with a decreasing kappa index is evident.

The emission frequency of gyromagnetic radiation is related to the electron gyrofrequency as follows:
\begin{equation}
    \nu=\frac{e}{2\pi m_e}sB=2.8\times10^{6}  {\; \rm Hz} \cdot sB\;[\rm G], 
    \label{equ:equ1}
\end{equation}
where $\nu$ is the emission frequency, $m_e$ is the electron mass, $s$ is the harmonic number and $B$ is the magnetic field. For the magnetic field at $\sim 700 \rm\; G$, peak emission frequency at $\sim 6 \rm\; GHz$ corresponds to the third harmonic frequency. The enhanced brightness temperature of the harmonic emission with a decreasing kappa index has been explained by \cite{2014ApJ...781...77F} in terms of an increase in optical depth of the harmonic layers and the enhancement of the effective temperature to greater than the kinetic temperature of the source plasma. Therefore, the decreasing kappa index may be best attributed to the proportional increase of the higher-energy, non-Maxwellian electrons in the emission source. However, owing to the limited angular resolution of EOVSA at low frequencies, the discussion on the optically thick part of the spectra may suffer from the beam dilution if the radio source is not well resolved. The results of the spectral fits may also be affected.

The energy spectra with the smallest kappa index for box A and box B shows about 10\% non-Maxwellian electrons but only less than 0.01\% $> 10$ keV electrons. Although the population of nonthermal electrons is insignificant in the energy spectra, we could not rule out their presence in this event because these nonthermal electrons can be quickly thermalized as they enter the dense footpoint region. The non-detection of a presumably weak nonthermal gyrosynchrotron source at the loop-top may be attributed to the limited dynamic range of EOVSA. In fact, as some researchers have argued that the kappa distribution can represent a solution of the transport equation for nonthermal electrons in the dense plasma \citep{2014ApJ...796..142B}, we argue that the transport and injection of nonthermal electrons into the footpoint region might account for the decrease, or hardening, of the kappa index during microwave enhancement. Alternatively, \cite{2015ApJ...802..122F} offered a different explanation for the decrease of the kappa index in solar flares, which is related to nonuniform cooling of the multi-temperature plasma. In our case, the decrease of the kappa index may also represent an increasing nonuniformity of the source due to thermal conduction.

\section{Discussions} \label{sec:discussions}
In the previous sections, we have discussed in detail how the enhanced microwave emission at the loop footpoints is connected with the EUV brightening of the ELP loops. Here, we would like to explain why the delayed microwave emission could result in delayed emission in high-temperature AIA passbands for ELP loops. As mentioned in section \ref{sec:ELP}, the AIA 131 \AA\ light curves for main flaring loops and ELP loops peak at 19:22 UT and 19:31 UT, respectively, indicating they undergo quite different heating processes. The main flaring loops are mainly energized by abrupt heating in region A during the impulsive phase, which is also illustrated by multiple radio bursts in Figure \ref{fig:figure7}a. The ELP loops, in contrast, have a much longer heating process, which extends to the gradual phase when the second stage of microwave enhancement emerges. Despite the timing, the ELP loops have a considerable energy input from region B, indicated by the evaporation flow in Figure \ref{fig:figure6}. Due to the longer duration and mild nature of the additional heating, the ELP loops reach the maximum temperature minutes after the main flaring loops. Although the role of the additional heating is often discussed in previous research, the duration of the heating is still under debate. In our case, we argue that the timescale of the heating is consistent with the delayed microwave emission enhancement, which lasts from 19:23 UT to 19:30 UT. Moreover, the AIA 131 \AA\ light curve of the ELP loops reaches its peak around 19:31 UT, meaning that the heating is likely ended at the time. However, we can not exclude that the additional heating may last much longer in other ELP events, even extending to the EUV late phase. The possibility could be tested by future observations.

\begin{figure*}
	\centering
	\includegraphics[width=\linewidth]{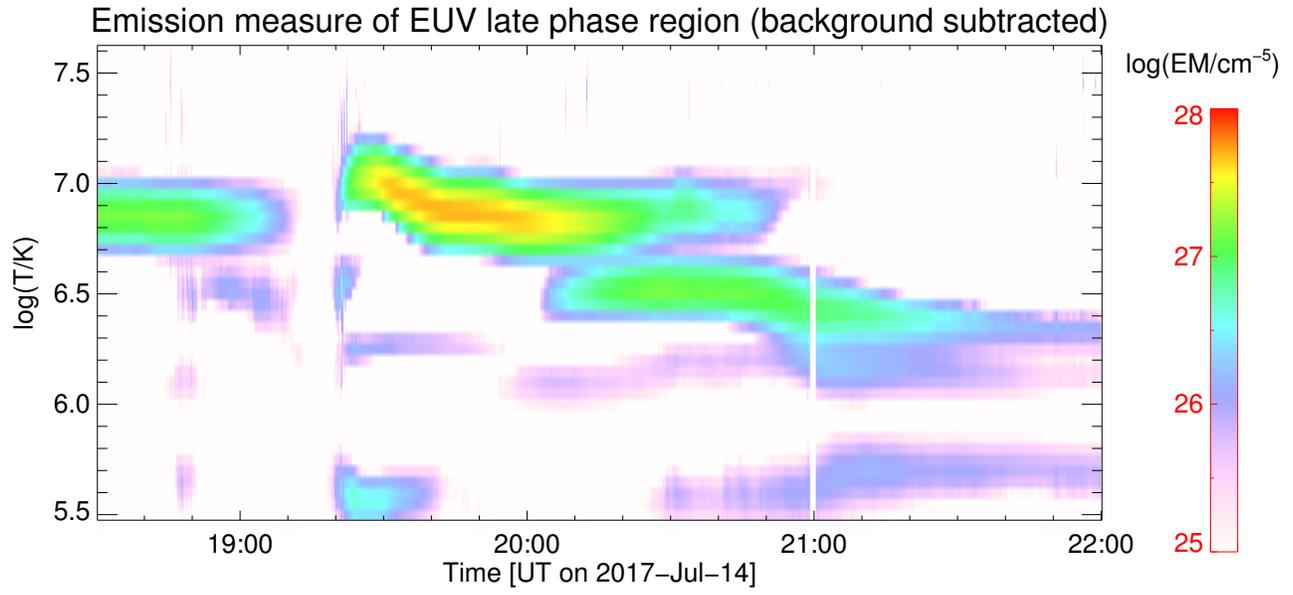}
	\caption{Evolution of the emission measure averaged over the ELP region after background subtraction. Background subtraction is to remove the irrelevant contribution in the line of sight. The background time is chosen as a time of EUV emission minimum (19:14:58 UT).\label{fig:figure9}}
\end{figure*}

\begin{figure*}
	\centering
	\includegraphics[width=\linewidth]{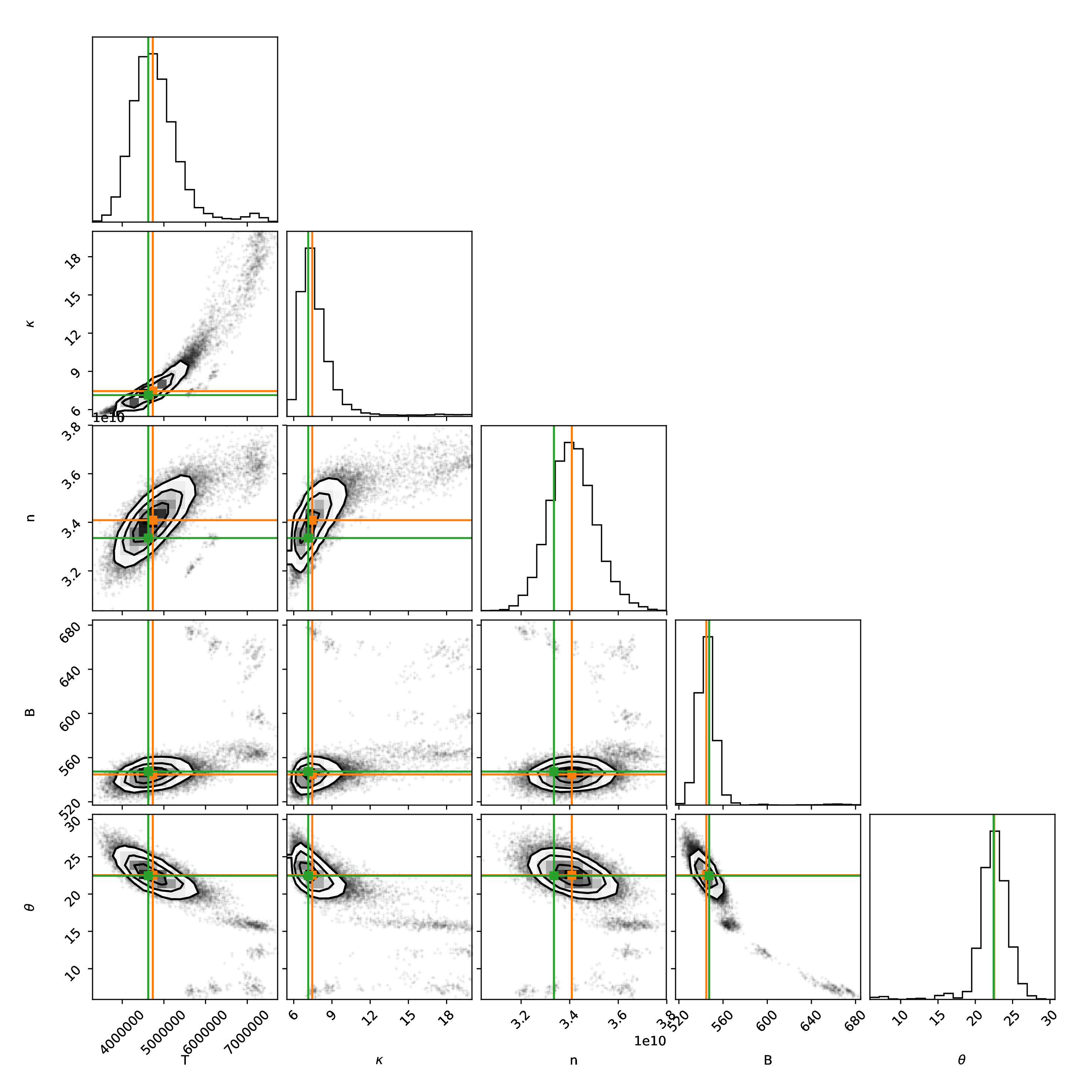}
	\caption{Markov chain Monte Carlo analysis for the spatially resolved microwave spectrum. The spectrum is taken from the box A at 19:21:13.500 UT in Figure \ref{fig:figure8}(1b). The green and orange lines represent the fitting results from the least-square-based fitting and MCMC analysis, respectively.\label{fig:figure10}}
\end{figure*}

\begin{figure*}
	\centering
	\includegraphics[width=\linewidth]{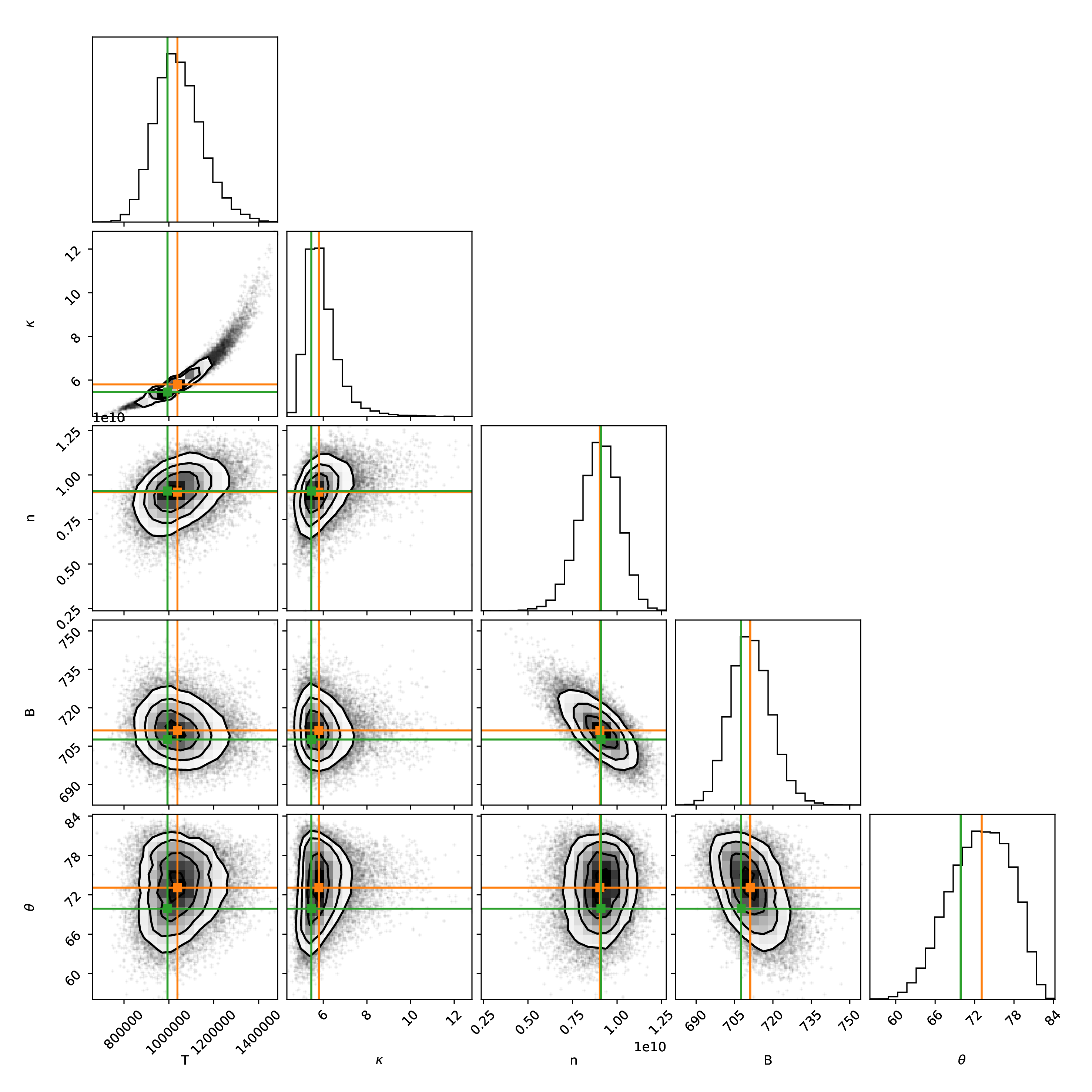}
	\caption{Markov chain Monte Carlo analysis for the spatially resolved microwave spectrum. The spectrum is taken from the box B at 19:23:52.500 UT in Figure \ref{fig:figure8}(2d). The green and orange lines represent the fitting results from the least-square-based fitting and MCMC analysis, respectively.\label{fig:figure11}}
\end{figure*}

For cooler AIA passbands, the emission peaks extending much later to the EUV late phase are determined by cooling instead of heating. In this case, the timescale of the ELP should correspond to the cooling timescale of the ELP loops. We derive the evolution of the emission measure of the ELP region (Figure \ref{fig:figure9}). The ELP loops cool from $10^{7.1}\rm\; K$ to $10^{6.4}\rm\;K$ in approximately two hours, with an average cooling rate of $1.4\times10^{3}\rm\;K/s$. Following \cite{1995ApJ...439.1034C}, the cooling time can be expressed as

\begin{equation}
	\tau_{cool}=0.0235\frac{L^{5/6}}{(Tn)^{1/6}}
\end{equation}

where $\tau_{cool}$ is the timescale of the cooling process in s, $L$ is the loop half length in cm, $T$ and $n$ are the initial temperature in K and density in $\rm cm^{-3}$, respectively. From AIA observations, we find $L\approx10^{10}\;\rm cm$. From the EM analysis, we obtain $T=13\;\rm MK$ and $n=6\times10^{9}\;\rm cm^{-3}$ (assuming an integration length of 3"). The cooling time is then estimated to be $\tau_{cool}\sim 129\rm\;min$, which is consistent with our observations. 

\section{Summary} \label{sec:conclusions}
We have studied a solar flare with a typical ELP and found direct evidence for the additional heating process that accounts for the ELP. The observed ELP is proceeded by a seven minute-long microwave enhancement. We find good correlations between the AIA 131 \AA\ derivative, UV light curve, and microwave light curve for the ELP. Signatures of chromospheric heating and evaporation are revealed by the UV and EUV images. The microwave spectra largely resemble the thermal gyroresonance emission. However, they deviate from the purely thermal case with a slightly rising spectral index in the optically thick regime, which we attribute to as the contribution from a non-Maxwellian component in the source electron distribution. We adopt the kappa distribution function as the model to fit the spectra and find a distinct decrease of the kappa index with an increasing microwave brightness temperature, indicating the possible rise of energetic electrons in the footpoint region.
Our analysis suggests that the heating of the ELP loops is likely driven by energy transfer from the flaring region to the loop footpoints and the subsequently induced chromospheric evaporation. We also distinguish that the time delay of high-temperature EUV emission from the ELP loops results from the gradual heating process and that of cooler EUV emission is mainly caused by the extended cooling process. 

Our work also brings new questions. First of all, is it a common phenomenon that the EUV enhancement in the flare late phase is preceded by a continuous microwave enhancement? We only examine one ELP event, so we could not draw a firm conclusion in this regard. Also, some researchers assume that additional heating could last through the whole ELP. In our case, the heating process appears to occur only at the beginning and is believed to end as the microwave enhanced emission disappears. However, we cannot exclude the possibility of a much longer heating process in other cases. Lastly, we do not fully confirm the cause of the minutes-long microwave enhancement, which represents the additional heating in the flare gradual phase. We offer two possibilities, injection of nonthermal electrons and multi-temperature plasma due to nonuniform heating. The true nature could be uncovered by future investigations.

\acknowledgments
This work is supported by National Key R\&D Program of China No. 2021YFA1600500 and NSFC grants 11825301 and 11790304. EOVSA operation is supported by NSF grant AST-1910354 and AGS-2130832 to NJIT. B.C. acknowledges support by NSF grant AGS-1654382 to NJIT. The SDO is a mission for NASA's Living With a Star (LWS) Program. The GOES is a joint effort of NASA and the National Oceanic and Atmospheric Administration (NOAA).





\end{document}